\documentstyle[11pt,paspconf,epsf]{article}
\begin{document}
\title{Reconstructing PSCz with a Generalized PIZA}
\author{Helen E.M. Valentine, Will Saunders and Andy Taylor}
\affil{Institute for Astronomy, University of Edinburgh, Royal Observatory, Blackford Hill, Edinburgh EH9 3HJ} 

\begin{abstract}
We present a generalized version of the Path Interchange Zel'dovich Approximation (PIZA; Croft and Gazta\~{n}aga 1997; hereafter CG97), which may be used to reconstruct density and velocity fields from realistic galaxy redshift surveys. The original PIZA does not include the possibility of using galaxies in redshift space with a selection function. We map galaxy positions from redshift to real space and from the Local Group rest frame to the CMB rest frame. To take account of the selection function, we minimize the mass-weighted action. We apply our new method to mock galaxy catalogues and find it offers an improvement in reconstructions over linear theory. Applying PIZA to the PSCz we can obtain real space density, initial density, and peculiar velocity fields, bulk velocities and the dipole.  Comparison of our reconstructed dipole with the CMB dipole gives $\beta\simeq 0.6$.
\end{abstract}
\section{Introduction to PIZA}
CG97 noted that with the Zel'dovich Approximation, the least action solution is that which minimizes the mean square particle displacement, \begin{equation} S=\sum_{\rm particles} m_i\xi_i^2\label{eq:one}\end{equation} where the $\xi_i$ are particle displacements. CG97 assume the galaxies have equal mass. The galaxies are assigned, at random, initial positions drawn from a homogeneous distribution.  These are represented by PIZA particles.

Applying PIZA is very simple. Two galaxies are picked at random and their particles are swapped. If S decreases, the swap is kept. This is repeated until S has been minimized.  CG97 found that $\sim$ 500 swaps per particle are needed.

\section{Problems with applying PIZA to redshift surveys}

\subsection{PIZA in Redshift Space}
PIZA minimizes real space galaxy displacements, but we have redshift space displacements. In linear theory, the two are related by (Taylor and Valentine 1999); 
\begin{equation}\xi_i={\cal P}^{-1}_{ij}\xi_j^s\end{equation}  where ${\cal P}^{-1}_{ij}\equiv\delta_{ij}^K-\frac{\beta}{1+\beta}\hat{s}_i\hat{s}_j$ is the inverse redshift space projection tensor.  The displacement squared is then
\begin{equation}\xi^2=\xi^{s^2}-(1-\frac{1}{(1+\beta)^2})\xi_r^{s^2}.\end{equation} 
\subsection{Frames of Reference}
PIZA is carried out in the CMB frame, but the survey redshifts are in the Local Group frame.
In the CMB Frame, 
\begin{equation} \xi^2=\xi^{s^2}-\xi_r^{s^2}+\frac{(\beta\xi_r^{LG}+\xi_r^s)^2}{(1+\beta)^2}\label{eq:lgr}   .\end{equation}
To find $\xi_r^{LG}$, a mock galaxy representing the Local Group is included in PIZA. 
\subsection{The Selection Function}
In a flux limited survey, the number density of objects decreases with increasing redshift.  We thus pick PIZA particles which obey the galaxy selection function, $\phi(r)$. To take account of galaxies below the flux limit, we assign each galaxy a mass of $\frac{1}{\phi(r)}$ and each particle a mass of $\frac{1}{n \phi(r)}$, where n is the particle-to-galaxy ratio. 
\subsection{Sky Coverage} 
There are regions on the sky, such as the Galactic Plane, that the survey does not cover.   We pick PIZA particles with the same sky coverage as the survey and neglect PIZA trajectories that cross the plane.
\section{Two-Step PIZA}
We have developed a two--step PIZA method to include these generalizations. Two steps are needed because there are two possibly conflicting conditions to satisfy -- the action must be minimized and the mass of each galaxy should be equal to the mass of its particles.
\par In the first step, each galaxy is assigned n particles. The particle masses are used in equation (\ref{eq:one}), and interchanges are carried out as usual. 
\par The second step attempts to improve on step one, by getting $M_{\rm galaxy}\sim \sum m_{\rm particles}$. A particle is taken from a galaxy with $M_{\rm galaxy}<\sum m_{\rm particles}$ (too many particles) and reassigned to a galaxy with $M_{\rm galaxy}>\sum m_{\rm particles}$ (too few). PIZA is then applied again, conserving the new particle to galaxy ratios. Generally this must be repeated 15 times.
\begin{figure}
\plotfiddle{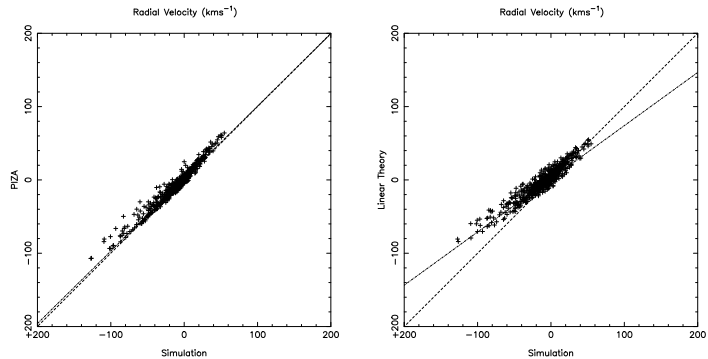}{4.2cm}{-90}{35}{35}{-160}{160}
\caption{Comparison of PIZA and Linear Theory Reconstructions. The left hand panel shows PIZA radial velocities compared with simulation, the right hand shows linear theory compared with simulation. Velocities were smoothed with a Gaussian of radius 20 Mpc.  The dotted line indicates the least-squares fit.  }
\label{fig:res2}
\end{figure}
\section{Tests on Simulations}
We have applied our new PIZA method to PSCz-like mock catalogues. We have tested the reconstruction of real space positions and radial peculiar velocities, and compared our reconstruction with linear theory. The mock catalogues and linear theory reconstructions were provided by Enzo Branchini (Branchini et al 1999).  Our conclusions from these tests include the following:

\par $\bullet$ The number of swaps needed is greater than that needed in the original PIZA -- approximately 4 times as many.
\par $\bullet$ Our reconstructions are generally better than linear theory (see Figure~\ref{fig:res2})
\par $\bullet$ The reconstructed Local Group velocity is not very sensitive to the assumed value of $\beta$.

\begin{figure}
\plotfiddle{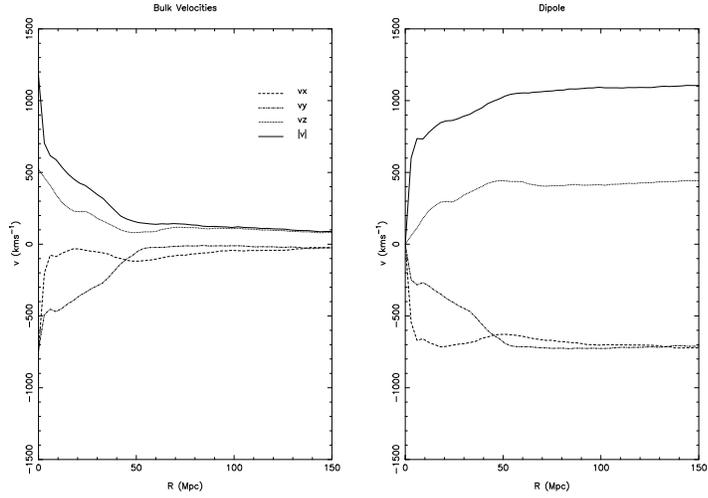}{5.5cm}{-90}{35}{35}{-140}{190}
\caption{The PSCz bulk flow and dipole.  The x-, y- and z- components and magnitude of the bulk flow and dipole are shown.  It can be seen that there is no contribution to the dipole after 100$h^{-1}$Mpc.  }
\label{fig:bulkf}
\end{figure}

\section{PIZA applied to PSCz}
 We have applied our new generalized PIZA method to the PSCz survey (Saunders et al 1995; also this volume), and present preliminary results here. Further results and analysis may be found in Valentine et al (1999).

\par  $\bullet$ The direction of the dipole was found to be $\sim 20^{\circ}$ away from the CMB dipole. Comparing the amplitude of these dipoles leads to a value of $\beta\simeq 0.6$.  Figure~\ref{fig:bulkf} shows the bulk flow and dipole.
\par $\bullet$ We have compared our bulk flows with those of Mk III and Branchini et al, and find good agreement with both, implying a value of $\beta\simeq 0.7$
\par
\begin{figure}
\plotfiddle{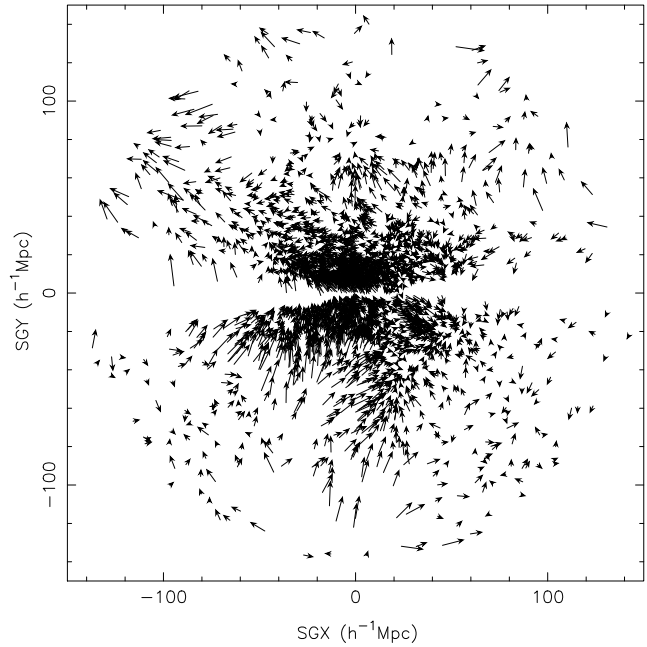}{7cm}{-90}{50}{50}{-200}{260}
\caption{A slice through the PSCz velocity field containing the Supergalactic Plane. The slice is 20 $h^{-1}$Mpc thick.  The arrows show the galaxy displacements, and the arrow heads are the real space galaxy positions.}
\label{fig:velf}
\end{figure}

 $\bullet$ The velocity field in the Supergalactic Plane is shown in Figure~\ref{fig:velf}.

\end{document}